\documentclass[journal=jctcce,manuscript=article]{achemso}

\usepackage{chemformula} 
\usepackage[T1]{fontenc} 
\usepackage{xcolor}
\usepackage{longtable}
\usepackage{tablefootnote}
\usepackage{amsmath} 
\usepackage{amssymb}
\usepackage{tcolorbox} 
\usepackage{enumerate} 
\usepackage{mathtools}
\usepackage{multirow}	

\newcommand{\onlinecite}[1]{\hspace{-2 ex} \citenum{#1}}
\newcommand{\modColor}{\color{black}}

\author{Chaoqun Zhang}
\email{czhan119@jhu.edu}
\affiliation{Department of Chemistry, The Johns Hopkins University, Baltimore, MD 21218, USA}
\author{Nicholas R. Hutzler}
\affiliation{Division of Physics, Mathematics, and Astronomy, California Institute of Technology, Pasadena, CA 91125, USA}
\author{Lan Cheng}
\affiliation{Department of Chemistry, The Johns Hopkins University, Baltimore, MD 21218, USA}

\title
  {Intensity-borrowing mechanisms pertinent to laser cooling of linear polyatomic molecules}

\begin{document}

\begin{abstract}
A study of the intensity-borrowing mechanisms important to optical cycling transitions in laser-coolable polyatomic molecules
arising from non-adiabatic coupling, contributions beyond the Franck-Condon approximation, and Fermi resonances
is reported.  
It has been shown to be necessary to include non-adiabatic coupling to obtain computational accuracy
that is sufficient to be useful for laser cooling of molecules.  
The predicted {\modColor vibronic} branching ratios using perturbation theory based on the non-adiabatic mechanisms have been demonstrated to
agree well with those obtained from variational discrete variable representation calculations for representative molecules including CaOH, SrOH, and YbOH. 
The electron-correlation and basis-set effects on the calculated 
transition properties, including the vibronic coupling constants, 
the spin-orbit coupling matrix elements, and the transition dipole moments, 
and on the calculated branching ratios have been thoroughly studied.
The vibronic branching ratios predicted using the present methodologies demonstrate 
that RaOH is a promising radioactive molecule candidate for laser cooling.
  
\end{abstract}

\section{Introduction}
Compared to atoms, cold molecules possess more intrinsic degrees of freedom and have emerged as a promising platform to cold chemistry \cite{Ospelkaus2010,Quemener2012,bohn2017cold}, quantum information science \cite{demille2002quantum,carr2009cold,wei2011entanglement,karra2016prospects,hudson2018dipolar,ni2018dipolar,yu2019scalable}, and precision measurement to the search for physics Beyond the Standard Model (BSM) \cite{zelevinsky2008precision,YbOH2017,Lim18,augenbraun2020molecular}. 
Direct laser cooling and trapping of molecules have recently been shown to be effective tools for preparing ultracold molecules \cite{Fitch2021Review,di2004laser,Stuhl2008,carr2009cold,Shuman2010,Hummon13,zhelyazkova2014laser,Barry14,Truppe17,Kozyryev2016,Anderegg17,Collopy2018,Williams18,baum20201d,SrOH2017,Ding2020,mitra2020direct,augenbraun2020molecular}.
An experimental laser-cooling scheme has been established for molecules~\cite{Fitch2021Review}: molecules are produced by laser ablation and then cooled with cryogenic buffer gas, slowed via radiation pressure from counter-propagating light, captured into a three-dimensional magneto-optical trap (3D MOT), and then transferred to a conservative trap, typically an optical dipole trap (ODT).
A variety of diatomic molecules have been slowed \cite{Shuman2010,Hummon13,zhelyazkova2014laser,Lim18} and loaded into 3D MOTs \cite{Barry14,Truppe17,Williams2017Characteristics,Anderegg17,Collopy2018,cheuk2018lambda,caldwell2019deep,Ding2020,langin2021polarization}.
The molecular temperature has been reduced to a few $\mu$K after sub-Doppler cooling \cite{caldwell2019deep,Ding2020,langin2021polarization,burau2023blue}.
While the extra internal degrees of freedom in polyatomic molecules are expected to provide unique applicability in quantum computing and precision measurements \cite{YbOH2017,yu2019scalable}, the laser cooling of polyatomic molecules is more challenging because of the more complicated vibrational and rotational structures~\cite{Augenbraun2023PolyLCReview}. 
Several polyatomic molecules have been successfully laser-cooled in one dimension, including SrOH, CaOH, YbOH, and CaOCH$_3$ \cite{SrOH2017,baum20201d,Augenbraun20a,mitra2020direct}. 
Very recently, the CaOH molecule has been directly laser-cooled in three dimensions, loaded into a 3D MOT \cite{vilas2022magneto}, sub-Doppler laser cooled to around 20 $\mu$K, and loaded into an optical dipole trap \cite{Hallas2022Optical}. We also mention that the polyatomic molecules
CH$_3$F \cite{zeppenfeld2012sisyphus, koller2022electric} and H$_2$CO \cite{prehn2016optoelectrical} have been cooled and trapped via optoelectrical cooling, {\modColor in which H$_2$CO molecule has been cooled to submillikelvin temperatures \cite{prehn2016optoelectrical}}, and CH$_3$ \cite{liu2017magnetic} has been magnetically trapped.

The laser-slowing step constitutes a major bottleneck for applying the laser-cooling scheme described above to polyatomic molecules.
To successfully slow molecules and load them into a 3D MOT, 
about $10^{4}$-$10^5$ photon scatterings per molecule are required \cite{di2004laser,Stuhl2008,Lasner2022vibronic}.
{\modColor This requires the leakage of an optical cycling to be smaller than $10^{-4}-10^{-5}$. In polyatomic molecules, the upper molecular state in an optical cycling can spontaneously decay to the vibrational states of the lower electronic state. All the decays with significant branching ratios, typically higher than $10^{-5}$ need to be addressed using repumping lasers to maintain the closedness of the optical cycling. This renders the laser systems for laser cooling of molecules not only technologically challenging, but also molecule specific and expensive.  } 
Therefore, the identification of promising laser-coolable molecules using both high-resolution spectroscopy and quantum-chemical calculations plays a key role in laser cooling of molecules.
The experimental measurements of {\modColor vibronic} branching ratios using high resolution laser spectroscopy \cite{Sheridan2007,Nakhate2019,Kozyryev2019,Paul2019,baum2021establishing,Nguyen2018,Mengesha2020,augenbraun2020molecular,lao2022laser,augenbraun2022high} have recently been combined with optical cycling to further enhance the sensitivity of the measurement \cite{zhang2021CaSrYb}.
Theoretical and computational tools are also being developed aiming to provide useful information about optical cycling transitions.
{\modColor Several criteria have been discussed concerning laser-coolability of molecules \cite{di2004laser,wells2011prospects,li2022theoretical}.}
Isaev and Berger have proposed that excitations localized at an atom in a polyatomic molecule leads to small structural differences between the ground and excited states \cite{Isaev2016}.
This gives nearly diagonal Franck-Condon factors with the branching ratio of the origin transition close to unity.
Many computational studies have been carried out 
following this criterion to look for promising candidates with quasidiagonal vibronic branching ratios  \cite{Li2019,Hao2019,Ivanov2019,ivanov2020toward,dickerson2022fully,osika2022fock,isaev2022radium}.
On the other hand, a critical experimental parameter is the number of repumping lasers required to address the branching ratios at the 99.99\% to 99.999\% level.
The knowledge of weak transitions with branching ratios above $10^{-5}$, including both weak symmetry-allowed transitions and the nominally symmetry-forbidden transitions, is therefore necessary in the design of deep-laser-cooling schemes.
{\modColor Take CaOH as an example. While the branching ratios for Ca-O stretching mode decays with the increase of vibrational quantum number as fast as those for Ca-F stretching mode in CaF, it has been found that it is necessary to repump a number of weak transitions for the bending modes that are nominally forbidden \cite{baum2021establishing}. In general in the laser-coolable linear triatomic molecules, relevant transitions can occur or be enhanced by means of "borrowing" intensities from strong transitions via spin-vibronic perturbations \cite{baum2021establishing,zhang2021CaSrYb}.
A thorough investigation of intensity-borrowing mechanisms with accurate treatments of spin-vibronic perturbations is required to assign mixed vibronic states and to provide insights into the weak transitions.} 

In a previous study \cite{zhang2021CaSrYb}, we have developed a computational scheme to calculate vibronic branching ratios by combining accurate potential energy surfaces and variational multi-state vibronic calculations. 
In this scheme, spin-vibronic perturbations are taken into account to enable an accurate description for weak transitions.
The computed vibronic energy levels for the $A^2\Pi$ and $X^2\Sigma^{+}$ states of CaOH and YbOH as well as the vibronic branching ratios for the $A^2\Pi\rightarrow X^2\Sigma^{+}$ transitions
agree very well with the experimental measurements, including for the nominally symmetry-forbidden transitions between states with bending angular momentum quanta differing by an odd number.
The calculations for the SrOH molecule have shown that it requires fewer repumping lasers to saturate 
vibronic branching to $99.999\%$ for the transitions from the vibrational ground state of the $A^2\Pi_{1/2}$ state to the vibrational states of the $X^2\Sigma^{+}$ state.
These results have facilitated the experimental measurements of the branching ratios and the design of a laser-cooling scheme for SrOH \cite{Lasner2022vibronic}.
{\modColor This paper features a first thorough study of the intensity-borrowing mechanisms for weak transition using perturbation theory (PT)}
together with variational discrete variable representation (DVR) calculations, in which the accuracy of calculated transition properties and branching ratios is systematically analyzed. 
In Sec. \ref{theorySec}, we recapitulate the quasidiabatic Hamiltonian with the spin-vibronic perturbations coupling the electronic excited states and discuss the intensity-borrowing mechanisms.
The computational details are summarized in Sec. \ref{compSec}.
A comparison between the vibronic branching ratios obtained from perturbational and variational calculations is presented in Sec. \ref{resultSec},
followed by benchmark calculations for the CaOH, SrOH, and YbOH molecules. Predictive calculations for RaOH are then presented to demonstrate that this molecule is a promising candidate for laser cooling.
Finally, we give a summary and an outlook in Sec. \ref{summarySec}.

\section{Theory}
\label{theorySec}

We focus our discussion on the metal-containing linear triatomic molecules of 
the type M-A-B, with the optical cycling center on the metal atom M.
In laser cooling of these molecules, the most commonly used optical cycle involves the transition from the vibrational ground state of a low-lying electronic excited state, e.g., the $A^2\Pi_{1/2}(000)$ state, to the vibrational states of the electronic ground state, i.e., $X^2\Sigma^{+}_{1/2}(000)$, $X^2\Sigma^{+}_{1/2}(100)$, $X^2\Sigma^{+}_{1/2}(02^00)$, and so on.
Here $(v_1 v_2^\ell v_3)$ denotes a vibrational state with $v_1$ quanta of the excitation in the M-A stretching mode, $v_2$ quanta of the bending excitation, and $v_3$ quanta of the A-B stretching excitation. The superscript $\ell$ denotes a vibrational angular momentum due to bending excitations.
{\modColor In metal monohydroxide, the excitations from the ground state to the $A^2\Pi$ and $B^2\Sigma^{+}$ states are mainly localized on the metal atom \cite{Isaev2016}. They correspond to the metal valence-shell $s\rightarrow p$ transitions.}
A number of small perturbations \cite{Fieldbook,Bernathbook, Hirotabook, HerzbergIII, baum20201d} {\modColor involving the $A^2\Pi$ and $B^2\Sigma^{+}$ states} play important roles {\modColor in the calculations of} 
the $A^2\Pi_{1/2}(000)\rightarrow X^2\Sigma^{+}_{1/2}(v_1 v_2^\ell v_3)$ transitions. These include the Renner-Teller (RT) effects of the $A^2\Pi$ states and the spin-orbit coupling (SOC) and the linear vibronic coupling (LVC) between the $A^2\Pi$ and $B^2\Sigma^{+}$ states. 
A full-fledged calculation of vibronic wave functions and branching ratios for linear triatomic molecules must include these effects \cite{carter1984variational,carter1990}. 
{\modColor It is necessary to account for the RT effects and SOC between $A^2\Pi_x$ and $A^2\Pi_y$ states to obtain an accurate description for the vibronic levels of the $A^2\Pi_{1/2}$ state. The SOC and LVC between $A^2\Pi_p$ ($p=x$ or $y$) and $B^2\Sigma^{+}$ states are responsible for the non-zero vibronic branching ratios for nominally symmetry-forbidden transitions.}

A powerful computational framework to treat non-adiabatic effects is the multi-state K{\"o}ppel-Domcke-Cederbaum (KDC) quasidiabatic Hamiltonian \cite{Koeppel84,sharma2021vibronically}.
We parametrize a six-state KDC Hamiltonian with the scalar quasidiabatic basis functions chosen as the $A^2\Pi_x(\alpha)$, $A^2\Pi_y(\alpha)$, $B^2\Sigma^{+}(\alpha)$, 
$A^2\Pi_x(\beta)$, $A^2\Pi_y(\beta)$, and $B^2\Sigma^{+}(\beta)$ states. Here $\alpha$ and $\beta$ refer to $S_z = 1/2$ and $-1/2$.
{\modColor The leading diabatic coupling between $A^2\Pi_x$ and $A^2\Pi_y$ states is quadratic in normal coordinates of molecular bending modes, while the leading diabatic coupling between $A^2\Pi_p$ ($p=x$ or $y$) and $B^2\Sigma^{+}$ states is linear in normal coordinates.}
This KDC Hamiltonian takes the form \cite{zhang2021CaSrYb}
\begin{eqnarray}
\begin{tiny}
    \left[ \begin{array}{cccccc}
        {E}^{\text{qd}}_{{A^2\Pi_x}}    & V^{\text{RT}}_{\text{xy}}+ih^{\text{SO}}_{\text{AA}} & V^{\text{LVC}}_{\text{xB}} & 0 & 0 & h^{\text{SO}}_{\text{AB}}\\ \\
        V^{\text{RT}}_{\text{yx}}-ih^{\text{SO}}_{\text{AA}} & {E}^{\text{qd}}_{{A^2\Pi_y}} & V^{\text{LVC}}_{\text{yB}} & 0 & 0 & ih^{\text{SO}}_{\text{AB}}\\ \\
        V^{\text{LVC}}_{\text{Bx}} & V^{\text{LVC}}_{\text{By}} & {E}^{\text{qd}}_{{B^2\Sigma^{+}}}  & -h^{\text{SO}}_{\text{BA}} & -ih^{\text{SO}}_{\text{BA}}& 0 \\ \\
           0 & 0  & -h^{\text{SO}}_{\text{AB}} & {E}^{\text{qd}}_{A^2\Pi_x} & V^{\text{RT}}_{\text{xy}}-ih^{\text{SO}}_{\text{AA}} & V^{\text{LVC}}_{\text{xB}} \\ \\
           0 & 0  & ih^{\text{SO}}_{\text{AB}} & V^{\text{RT}}_{\text{yx}}+ih^{\text{SO}}_{\text{AA}} & {E}^{\text{qd}}_{A^2\Pi_y}  & V^{\text{LVC}}_{\text{yB}} \\ \\
          h^{\text{SO}}_{\text{BA}}   & -ih^{\text{SO}}_{\text{BA}}  & 0 & V^{\text{LVC}}_{\text{Bx}} & V^{\text{LVC}}_{\text{By}}  & {E}^{\text{qd}}_{B^2\Sigma^{+}}     
  \end{array} \right].
\end{tiny}
\label{6sH}
\end{eqnarray}
${E}^{\text{qd}}_{A^2\Pi_x}$, ${E}^{\text{qd}}_{A^2\Pi_y}$, and ${E}^{\text{qd}}_{B^2\Sigma^{+}}$ 
denote the potential energies of these quasidiabatic states.
$V^{\text{RT}}$'s represent RT coupling, 
$V^{\text{LVC}}$'s represent LVC, and $h^{\text{SO}}$'s 
denote SOC matrices.

We have obtained the adiabatic potential energy surfaces for $B^2\Sigma^{+}$ and the two scalar adiabatic $A^2\Pi$ states, i,e., the $A^2\Pi_+$ and $A^2\Pi_-$ states, from quantum-chemical calculations. 
To treat the RT effects, the $A^2\Pi_+$ and $A^2\Pi_-$ states have been transformed into the quasidiabatic $A^2\Pi_x$ and $A^2\Pi_y$ states with a second-order vibronic coupling between them \cite{zhang2021CaSrYb}.
Next, we have included the LVC and SOC between the scalar electronic states and obtained the Eqn. (\ref{6sH}). 
The vibronic coupling matrix elements are given as terms linear in the normal coordinates $Q_x$ or $Q_y$. 
For example, the linear vibronic coupling between $B^2\Sigma^{+}$ and $A^2\Pi_x$ is given by
\begin{eqnarray}
  \label{lambda}
  V_{\text{Bx}}^{\text{LVC}}=V_{\text{xB}}^{\text{LVC}}=\lambda Q_x,
\end{eqnarray}
in which $\lambda$ is the corresponding linear diabatic coupling constant \cite{Ichino09}.

Weak transitions
, including the nominally symmetry-forbidden $\left|\Delta v_2\right| = 1, 3, 5, \cdots$ transitions, are 
closely relevant to the design of deep-laser-cooling schemes \cite{baum2021establishing, Augenbraun2021YbOCH3}. 
For the $\left|\Delta v_2\right| = 1$ transitions,
the branching ratios of the $A^2\Pi_{1/2}(000) \rightarrow X^2\Sigma^{+}_{1/2}(010)$ transitions range from $10^{-4}$ to $10^{-3}$ in CaOH, SrOH, and YbOH \cite{zhang2021CaSrYb,Lasner2022vibronic}.
The $B^2\Sigma^{+}_{1/2}(000)\rightarrow X^2\Sigma^{+}_{1/2}(010)$ transitions are stronger.
In the SrOH molecule, the branching ratio \cite{Lasner2022vibronic} of the $B^2\Sigma^{+}_{1/2}(000)\rightarrow X^2\Sigma^{+}_{1/2}(010)$ transition is 0.2\% and about one order of magnitude larger than the value of 0.04\% for the $A^2\Pi_{1/2}(000)\rightarrow X^2\Sigma^{+}_{1/2}(010)$ transition.
While a variational calculation provides accurate results, 
a perturbational analysis of the intensity-borrowing mechanisms can provide valuable insights into the dependence of the branching ratios to vibronic energy levels and transition properties.
In practice, the perturbational scheme enables estimate of intensities for the weak symmetry-forbidden transition using vibrational energy levels and transition properties, without a full-fledged variational vibrational calculation.
This is especially useful for treating larger molecules, for which full-dimensional potential energy surfaces and variational vibrational calculations are beyond the capability of present computational resources.

{\modColor In the study using perturbation theory, it is simpler and more accurate}
to work with the diabatic electronic basis functions chosen as $A^2\Pi_{1/2}$, $A^2\Pi_{3/2}$, and $B^2\Sigma^{+}_{1/2}$ {\modColor instead of $A^2\Pi_x$ and $A^2\Pi_y$ used in the DVR calculations} to analyze the intensity-borrowing mechanisms. 
In this representation, {\modColor the SOC within the $A^2\Pi$ states enter as diagonal energy corrections for $A^2\Pi_{1/2}$ and $A^2\Pi_{3/2}$.}
The remaining perturbations are the spin-orbit coupling and the LVC between $A^2\Pi_{1/2}$ and $B^2\Sigma^{+}_{1/2}$ and {\modColor the LVC} between $A^2\Pi_{3/2}$ and $B^2\Sigma^{+}_{1/2}$. 
These couplings contribute important intensity-borrowing mechanisms that will be discussed in Section \ref{borrow nac}. 
Apart from vibronic coupling, the contributions from the derivatives of transition dipole moments can also contribute intensities to the forbidden transitions. This will be presented in Section \ref{borrow tdmd}. 
Furthermore, we will discuss Fermi resonances that redistribute intensities among symmetry-allowed transitions and are responsible for the intensity borrowing for weak symmetry-allowed transitions.

\subsection{Intensity-borrowing mechanisms via non-adiabatic coupling}
\label{borrow nac}
We have briefly discussed the intensity-borrowing mechanisms via non-adiabatic coupling for the symmetry-forbidden $A^2\Pi_{1/2}(000)\rightarrow X^2\Sigma^{+}_{1/2}(010)$ transition in Ref.~ \onlinecite{zhang2021CaSrYb}. 
Here we give a thorough discussion for this transition and extend the formulation to the $A^2\Pi_{3/2}(000)\rightarrow X^2\Sigma^{+}_{1/2}(010)$ and $B^2\Sigma^{+}_{1/2}(000)\rightarrow X^2\Sigma^{+}_{1/2}(010)$ transitions.

We refer to a first-order mechanism as the ``direct vibronic coupling'' (DVC) mechanism. 
The linear vibronic coupling can mix a $B^2\Sigma^{+}_{1/2}(010)$ component directly into the $A^2\Pi_{1/2}(000)$ wave function. 
The dipole strength, i.e., the square of the norm of the transition dipole moment, borrowed through the DVC mechanism can be evaluate using perturbation theory as
\begin{equation}
    \label{ax_dvc_original}
    |\Vec{d}_{DVC}|^2 = \frac{|\langle A^2\Pi_{1/2}(000)| \hat{h}^{\text{LVC}} | B^2\Sigma^{+}_{1/2}(010) \rangle |^2}{(\Delta E_{AB} + \omega^{B}(010))^2} |h^{\text{dip}}_{\text{BX}}|^2,
\end{equation}
in which $h^{\text{dip}}_{\text{BX}}$
is the electronic transition dipole moment between the $X^2\Sigma^{+}_{1/2}$ and $B^2\Sigma^{+}_{1/2}$ states, $\Delta E_{AB}$ is the energy splitting between the $A^2\Pi_{1/2}$ and $B^2\Sigma^{+}_{1/2}$ states, and $\omega^B(010)$ is the level position of $B^2\Sigma^{+}_{1/2}(010)$ relative to $B^2\Sigma^{+}_{1/2}(000)$.
    
In a ``spin-orbit-vibronic coupling'' (SOVC) mechanism, 
the $A^2\Pi_{1/2}(000)$ state is mixed with the $B^2\Sigma^{+}_{1/2}(000)$ state via SOC
and the $B^2\Sigma^{+}_{1/2}(000)$ state is then coupled with $A^2\Pi_{3/2}(010)$ via LVC. 
The dipole strength obtained from the SOVC mechanism can be calculated using perturbation theory as
\begin{equation}
   \label{ax_sovc_original}
   |\Vec{d}_{SOVC}|^2 = \frac{|\langle A^2\Pi_{1/2}(000)| \hat{h}^{\text{SO}} | B^2\Sigma^{+}_{1/2}(000) \rangle \langle B^2\Sigma^{+}_{1/2}(000)| \hat{h}^{\text{LVC}} | A^2\Pi_{3/2}(010) \rangle |^2}{(\Delta E_{AB})^2(\Delta E_{AA} + \omega^{A}(010))^2} |h^{\text{dip}}_{\text{AX}}|^2,
\end{equation}
in which $h^{\text{dip}}_{\text{AX}}$
is the electronic transition dipole moment between the $X^2\Sigma^{+}_{1/2}$ and $A^2\Pi_{1/2}$ states, $\Delta E_{AA}$ is the energy splitting between the $A^2\Pi_{1/2}$ and $A^2\Pi_{3/2}$ states, and $\omega^A(010)$ is the level position of $A^2\Pi_{1/2}(010)$ relative to $A^2\Pi_{1/2}(000)$.

\begin{figure}
    \centering
    \includegraphics[width=8cm]{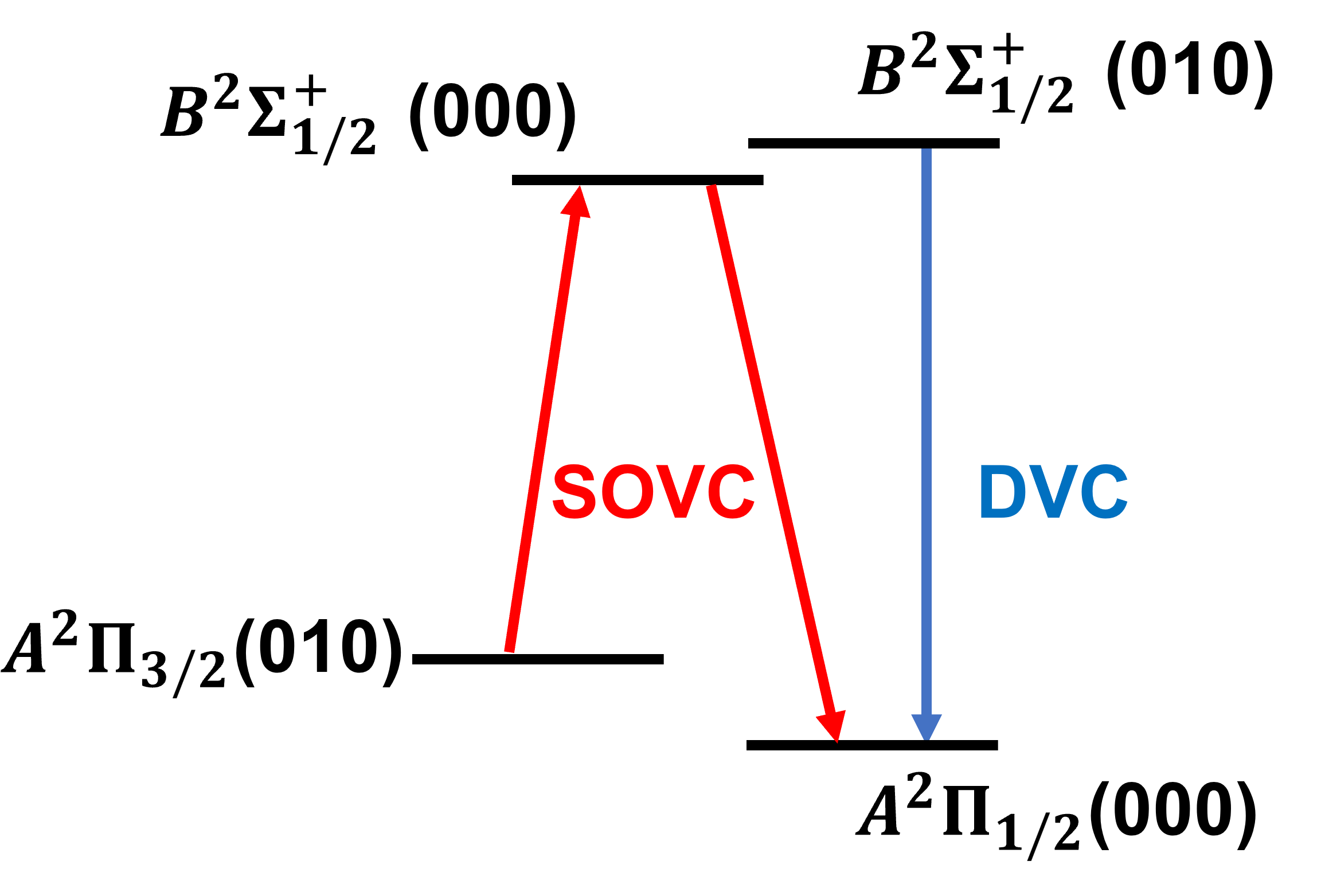}
    \caption{The mixing of $A^2\Pi_{3/2}(010)$ component into the $A^2\Pi_{1/2}(000)$ wave function via SOVC mechanism and the mixing of $B^2\Sigma^{+}_{1/2}(010)$ component into the $A^2\Pi_{1/2}(000)$ wave function via DVC mechanisms.
    }
    \label{ibm_A1}
\end{figure}

The mixing of $A^2\Pi_{3/2}(010)$ and $B^2\Sigma^{+}_{1/2}(010)$ components into the $A^2\Pi_{1/2}(000)$ wave function arising from the DVC and SOVC mechanisms is illustrated in Fig. \ref{ibm_A1}.
{\modColor  In terms of intensity-borrowing, the $A^2\Pi_{1/2}(000)\rightarrow X^2\Sigma^{+}_{1/2}(010)$ transition can borrow intensity from the $A^2\Pi_{3/2}(010)\rightarrow X^2\Sigma^{+}_{1/2}(010)$ and the $B^2\Sigma^{+}_{1/2}(010)\rightarrow X^2\Sigma^{+}_{1/2}(010)$ transitions via SOVC and DVC mechanisms, respectively, as indicated by the arrows in Fig. \ref{ibm_A1}.}
We note that the Eqns. (\ref{ax_dvc_original}) and (\ref{ax_sovc_original}) resemble the Eqns. (8) and (9) in Ref.~ \onlinecite{baum2021establishing} with the difference lying in the choice of basis functions in the perturbation theory. 
{\modColor The $A^2\Pi_{1/2}$ and $A^2\Pi_{3/2}$ states have been used as the basis functions in our work, while the scalar $A^2\Pi_x$ and $A^2\Pi_y$ states have been used in Ref.~ \onlinecite{baum2021establishing}.} This leads to a major difference {\modColor that the spin-orbit corrections to the energies of the $A^2\Pi_{1/2}$ and $A^2\Pi_{3/2}$ states can be included in the denominators in the present analysis.}
Since the magnitude of spin-orbit splittings between $A^2\Pi_{1/2}$ and $A^2\Pi_{3/2}$ states become comparable or larger than the vibrational spacing for molecules containing heavy atoms, it is important to use a denominator with the inclusion of the spin-orbit splitting $\Delta E_{AA}$. 
Interestingly, the DVC mechanism still dominates for the YbOH molecule, which exhibits strong spin-orbit coupling. 
The reason is that the large magnitude of the spin-orbit coupling matrix element $\langle A^2\Pi_{1/2}(000)| \hat{h}^{\text{SO}} | B^2\Sigma^{+}_{1/2}(000) \rangle$ in the numerator is offset by the equally large magnitude of the spin-orbit splitting $\Delta E_{AA}$ in the denominator \cite{zhang2021CaSrYb}.

Since the spin-orbit coupling matrix elements vary slowly with respect to the geometrical displacement, we use the electronic spin-orbit matrix elements for $\langle A^2\Pi_{1/2}(000)| \hat{h}^{\text{SO}} | B^2\Sigma^{+}_{1/2}(000) \rangle$.
Within the harmonic approximation assuming that the $A^2\Pi_x$, $A^2\Pi_y$ and $B^2\Sigma^{+}_{1/2}$ states share the same set of vibrational wave functions,
the linear vibronic coupling matrix elements in Eqns. (\ref{ax_dvc_original}) and (\ref{ax_sovc_original}) take simple forms, i.e.,
\begin{equation}
    \left|\langle B^2\Sigma^{+}_{1/2}(000)| \hat{h}^{\text{LVC}} | A^2\Pi_{x}(010) \rangle\right| \approx \frac{\lambda}{\sqrt{2}},
\end{equation}
{\modColor where $\lambda$ is the linear quasidiabatic coupling constant defined in Eqn. (\ref{lambda}).}
The intensities borrowed from these two mechanisms are proportional to the corresponding dipole strengths, leading to
\begin{equation}
\label{ibm_dvc_A1}
    I_{DVC}\left[A^2\Pi_{1/2}(000)\rightarrow X^2\Sigma^{+}_{1/2}(010)\right] \propto \frac{|\lambda|^2}{2(\Delta E_{AB} + \omega^{B}(010))^2} |h^{\text{dip}}_{\text{BX}}|^2
\end{equation}
and
\begin{equation}
    I_{SOVC}\left[A^2\Pi_{1/2}(000)\rightarrow X^2\Sigma^{+}_{1/2}(010)\right] \propto \frac{|h^{\text{SO}}_{\text{AB}}|^2  |\lambda|^2}{(\Delta E_{AB})^2(\Delta E_{AA} + \omega^{A}(010))^2} |h^{\text{dip}}_{\text{AX}}|^2,
\end{equation}
respectively.
Similarly, the corresponding intensities of the $A^2\Pi_{3/2}(000)\rightarrow X^2\Sigma^{+}_{1/2}(010)$ and $B^2\Sigma^{+}_{1/2}(000)\rightarrow X^2\Sigma^{+}_{1/2}(010)$ transitions can be evaluated as
\begin{equation}
    I_{DVC}\left[A^2\Pi_{3/2}(000)\rightarrow X^2\Sigma^{+}_{1/2}(010)\right] \propto \frac{|\lambda|^2}{2(\Delta E_{AB} - \Delta E_{AA} + \omega^{B}(010))^2} |h^{\text{dip}}_{\text{BX}}|^2,
\end{equation}
\begin{equation}
    I_{SOVC}\left[A^2\Pi_{3/2}(000)\rightarrow X^2\Sigma^{+}_{1/2}(010)\right] \propto \frac{|h^{\text{SO}}_{\text{AB}}|^2  |\lambda|^2}{(\Delta E_{AB} - \Delta E_{AA} + \omega^{B}(010))^2(\Delta E_{AA} - \omega^{A}(010))^2} |h^{\text{dip}}_{\text{AX}}|^2,
\end{equation}
and
\begin{equation}
\label{ibm_dvc_B}
\begin{aligned}
    I_{DVC}\left[B^2\Sigma^{+}_{1/2}(000)\rightarrow X^2\Sigma^{+}_{1/2}(010)\right] 
    & \propto\frac{|\lambda|^2}{2(\Delta E_{AB} - \omega^{A}(010))^2} |h^{\text{dip}}_{\text{AX}}|^2\\
    & +\frac{|\lambda|^2}{2(\Delta E_{AB} - \Delta E_{AA} - \omega^{A}(010))^2} |h^{\text{dip}}_{\text{AX}}|^2.
\end{aligned}
\end{equation}
Note that the $B^2\Sigma^{+}_{1/2}(000)\rightarrow X^2\Sigma^{+}_{1/2}(010)$ transition does not borrow intensities via the SOVC mechanism. Instead, it borrows intensities from both $A^2\Pi_{1/2}(010)$ and $A^2\Pi_{3/2}(010)$ states via the DVC mechanism.

\subsection{The transition dipole moment derivatives mechanism beyond the Franck-Condon approximation}
\label{borrow tdmd}
The Franck-Condon approximation neglects the geometrical dependence of the electronic transition dipole moments.
We refer to the intensity borrowing due to the geometrical dependence of the transition dipole moments as the ``transition dipole moment derivatives'' (TDMD) mechanism.  
Of particular interest here is that the derivatives of the electronic transition dipole moment between the $A^2\Pi_x$ (or $A^2\Pi_y$) and $X^2\Sigma^{+}_{1/2}$ states contribute to the nominally symmetry-forbidden $A^2\Pi_{1/2}(000)\rightarrow X^2\Sigma^{+}_{1/2}(010)$ transition. 
This leads to a first-order corrections to the intensities proportional to
\begin{equation}
    \frac{\partial h^{\text{dip}}_{\text{AX},z}}{\partial Q_x}  \langle (000) | Q_x | (010)\rangle
\end{equation}
and
\begin{equation}
    \frac{\partial h^{\text{dip}}_{\text{AX},z}}{\partial Q_y}  \langle (000) | Q_y | (010)\rangle,
\end{equation}
in which $h^{\text{dip}}_{\text{AX},z}$
is the z-component of the electronic transition dipole moment $h^{\text{dip}}_{\text{AX}}$, $Q_x$ and $Q_y$ are the normal coordinates for the molecular bending along $x$ and $y$ directions, respectively. 

\subsection{Intensity borrowing from Fermi resonance}

Fermi resonances mix vibronic states that are close in energy \cite{fermi1931ramaneffekt,hougen1962fermi}. 
Some weak transitions can be enhanced by Fermi resonances, when they are nearly degenerate to a strong transition.
As a result, Fermi resonance redistributes the intensities between these transitions and induce more branching ratios non-negligible in laser-cooling experiments.
This mechanism is particularly important in molecules of the M-A-B type when the vibrational energy of the $(020)$ state are close to that of the $(100)$ state. Similar consideration applies to the $(120)$, $(210)$, and $(040)$ states. 
In Section \ref{fermiResults}, we present calculations to demonstrate the effects of Fermi resonances on the intensities of these states.

\section{Computational details}
\label{compSec}

The calculations of the adiabatic potential energy surfaces (PESs) and the coupling parameters in the KDC Hamiltonian have been performed using the CFOUR program \cite{CFOUR1,CFOUR2,Stanton99,cheng11b}. 
The potential energy surfaces of the electronic ground and excited states of CaOH, SrOH, and YbOH have been obtained using the equation-of-motion coupled-cluster singles and doubles for electron attachment (EOMEA-CCSD) method \cite{EOMCC1Stanton1993,EOMEA_Nooijen95} together with correlation-consistent basis sets \cite{dunning1989gaussian,koput2002ab,de2001parallel,hill2017gaussian,lu2016correlation}.
We have used the cc-pCVXZ (X=T,Q) basis sets for the Ca atom, cc-pwCVTZ and cc-pVTZ basis sets recontracted for the spin-free exact two-component theory in its one-electron variant (SFX2C-1e) \cite{sfx2c1e2001dyall,Liu2009} for the Sr and Yb atoms, respectively.
We denote the combination of the cc-p(wC)VXZ 
basis set for a metal atom and the cc-pVXZ basis sets for the O and H atoms as the ``XZ'' basis.
More details about the basis sets, the number of frozen-core orbitals, and the choice of the grid points for the construction of the PESs in the calculations for CaOH, SrOH, and YbOH have been documented in Ref.~~\onlinecite{zhang2021CaSrYb}.

For the RaOH molecule, the PESs have been calculated at the SFX2C-1e-EOMEA-CCSD level using the uncontracted ANO-RCC basis set \cite{ANORCC1,ANORCC_main} for Ra and the uncontracted cc-pVTZ basis sets for O and H. 
The lowest 35 core orbitals and the virtual orbitals above 200 hartree have been kept frozen
in the coupled-cluster calculations.
The calculations of RaOH have been carried out for 2187 grid points on the potential energy surface of each state (11 evenly spaced grid points covering the range [3.8979 bohr, 4.8979 bohr] for the Ra-O bond length, 11 evenly spaced grid points covering the range [1.6004 bohr, 2.1004 bohr] for the H-O bond length, and 18 evenly spaced grid points covering the range [$95^{\circ}$, $180^{\circ}$] for the Ra-O-H angle). 
The analytical potential energy functions have been obtained by fitting the \textit{ab initio} energies into sixth-order polynomials in terms of three internal coordinate displacements from the equilibrium structure of the $X^2\Sigma^{+}_{1/2}$ state, i.e., $E(r_1,r_2,\theta) = \sum_{i,j,k}A_{ijk}(1/(i!j!k!))(r_1-4.2979)^i(r_2-1.8004)^j(\theta-180^{\circ})^k$, where $r_1$, $r_2$, and $\theta$ denote the Ra-O bond length, O-H bond length, and Ra-O-H angle, respectively.

The scalar-relativistic effects have been treated using the SFX2C-1e method in the calculations of SrOH, YbOH, and RaOH. The spin-orbit matrix elements and transition dipole moments have been calculated using the EOMEA-CCSD expectation-value formulation \cite{Klein08,cheng18a}. 
The molecular effective one-electron spin-orbit integrals have been calculated using the exact two-component approach
with atomic mean-field spin-orbit integrals (the X2CAMF scheme) constructed for perturbative treatments of spin-orbit coupling \cite{cheng18a,zhang2020performance}.
The linear diabatic vibronic coupling constants $\lambda$'s have been evaluated using the analytic scheme developed in Ref.~~\onlinecite{Ichino09}.
For the RaOH molecule, these coupling matrix elements have been calculated using the same basis sets as those used in the calculation of PESs with all the electrons correlated.

The vibronic energy levels and branching ratios have been obtained using the discrete variable representation method \cite{light1985generalized,Colbert1992,light2000discrete} in the normal-coordinate representation. 
The conventional rectangular kinetic energy formulae developed in Ref.~ \onlinecite{Colbert1992} have been used to evaluate the nuclear kinetic energy matrix elements.
The vibrational states of the $X^2\Sigma^{+}_{1/2}$ state have been obtained from single-state calculations, since the ground state is well separated from the other electronic states.
The six-state quasidiabatic Hamiltonian defined in Eqn. (\ref{6sH}) have been used for the calculations of vibronic states of the $A^2\Pi_{1/2}$, $A^2\Pi_{3/2}$, and $B^2\Sigma^{+}_{1/2}$ states.
The branching ratio for a transition from an upper state to a vibrational state $\mu$ of the ground electronic state,
i.e., the $X^2\Sigma^{+}_{1/2}(\mu)$ state, have been calculated using the $(\Delta E)_\mu^3$-weighted formula
\begin{equation}
    b_{\mu}=\frac{(\Delta E)_\mu^3|\Vec{d}_{\mu}|^2}{\sum_{\nu}(\Delta E)_\nu^3|\Vec{d}_{\nu}|^2},
\end{equation}
where $(\Delta E)_\mu$ and $\Vec{d}_{\mu}$ refer to the transition energies and transition dipole moment vectors between the upper vibronic state and the $X^2\Sigma^{+}_{1/2}(\mu)$ states.

\section{Results and discussion}
\label{resultSec}
\subsection{The importance of anharmonic and non-adiabatic contributions}
\begin{figure}
    \centering
    \includegraphics[width=10cm]{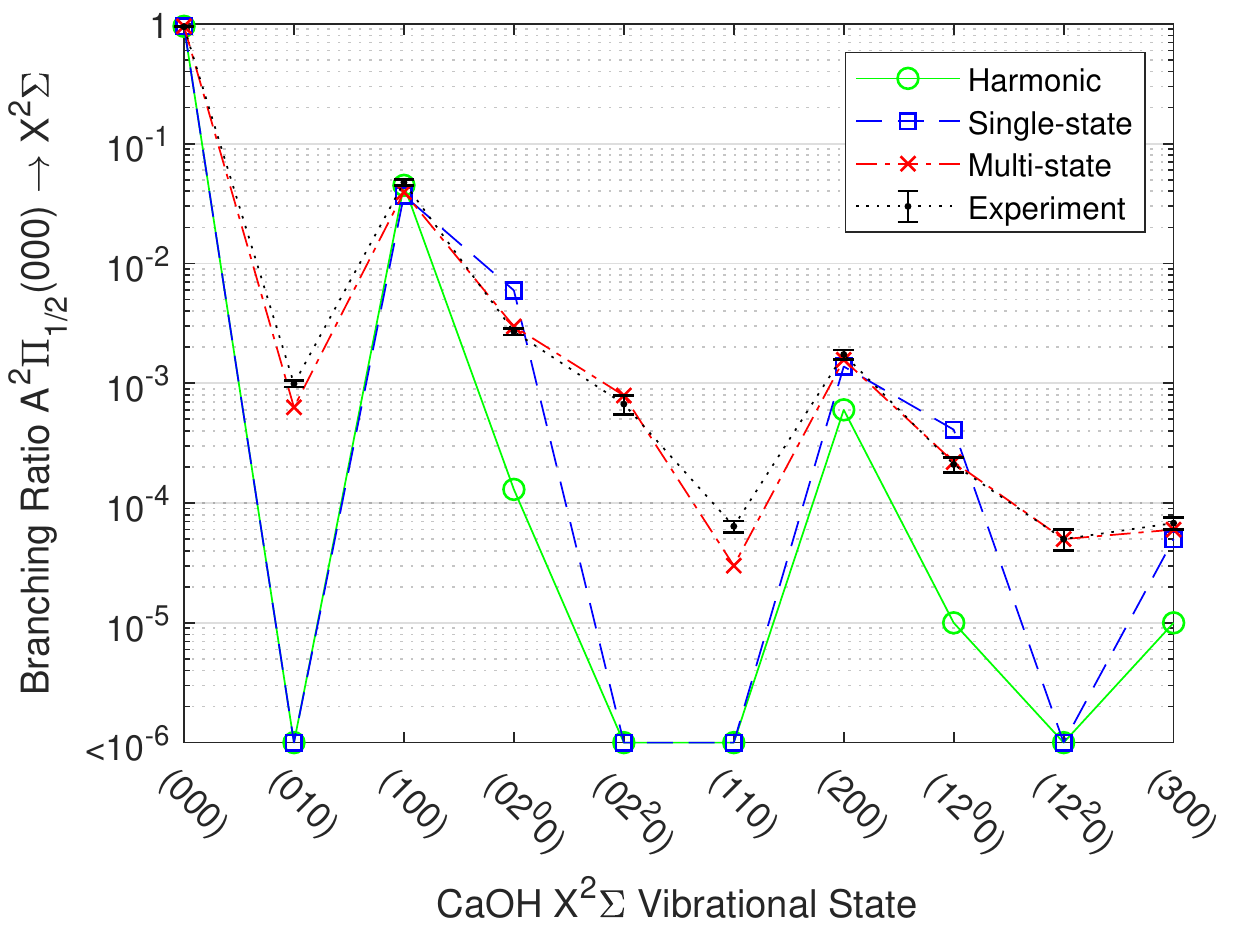}
    \caption{Comparison between the harmonic approximation, single-state and multi-state DVR calculations, and experimental values~\cite{zhang2021CaSrYb} for the branching ratios of $A^2\Pi_{1/2}(000)$ to vibrational levels of $X^2\Sigma^{+}$ in the CaOH molecule.
    }
    \label{comparison}
\end{figure}

We first present a comparison between the harmonic approximation, a single-state DVR calculation, a multi-state DVR calculation, and the experimental measurement for the branching ratios of the strongest ten $A^2\Pi_{1/2}(000)\rightarrow X^2\Sigma^{+}_{1/2}(v_1 v_2 v_3)$ vibronic transitions in CaOH.
Compared with the harmonic approximation, the single-state DVR calculation takes the anharmonic contributions into account.
The differences between the multi- and single-state DVR calculations include the non-adiabatic effects and the spin-orbit coupling contributions.
The deep laser cooling of CaOH has used nine repumping lasers to include all these transitions in the optical cycle to ensure the closedness of the optical cycle \cite{vilas2022magneto}. 
In order for a computational method to be useful for facilitating laser-cooling experiments, it is necessary to obtain accurate branching ratios for all these transitions.
As shown in Figure \ref{comparison} and Table \ref{harmonic}, the harmonic approximation and the single-state DVR calculation predicts three and five of these transitions to have branching ratios above $10^{-5}$, which corresponds to the use of 3 and 5 repumping lasers, respectively.
Only the multi-state DVR calculation is capable of accurately describing all nine transitions. It thus is absolutely necessary to go beyond the harmonic approximation and to include both the anharmonic contributions and non-adiabatic effects.

The harmonic approximation gives accurate branching ratios for the $A^2\Pi_{1/2}(000)\rightarrow X^2\Sigma^{+}_{1/2}(000)$ transition.
However, the branching ratios decay too quickly as $v_1$ (the Ca-O stretching mode) increases, and the branching ratios of the $A^2\Pi_{1/2}(000)\rightarrow X^2\Sigma^{+}_{1/2}(200)$ and $X^2\Sigma^{+}_{1/2}(300)$ transitions are underestimated.
For the molecular bending mode, since the PESs along the bond angles are flat and the anharmonic effects are more significant, the branching ratio of $A^2\Pi_{1/2}(000)\rightarrow X^2\Sigma^{+}_{1/2}(020)$ obtained from the harmonic approximation is one order of magnitude smaller than the measured value.
The calculated branching ratios obtained from single-state and multi-state DVR calculations agree very well with each other for the origin transition and the $A^2\Pi_{1/2}(000)\rightarrow X^2\Sigma^{+}_{1/2}(v_1 00)$ transitions for $v_1=1,2,3$.
However, the single-state calculation without taking into account the non-adiabatic effects misses all the nominally symmetry-forbidden transitions.
The inclusion of the linear vibronic coupling and spin-orbit coupling in the multi-state calculation enables capturing the transitions with $\left|\Delta v_2\right| = 1, 3, 5, \cdots$. In addition, the Renner-Teller effects redistribute the intensities among the transitions to the $X^2\Sigma^{+}_{1/2}(v_1 2^00)$ and $X^2\Sigma^{+}_{1/2}(v_1 2^20)$ states with $v_1 = 1$ and $2$.

\subsection{Numerical assessment of the intensity-borrowing mechanisms}
\subsubsection{DVC and SOVC mechanisms}
In the perturbation theory, we assume the same vibrational wave functions for the $X^2\Sigma^{+}_{1/2}$, $A^2\Pi_{1/2}$, $A^2\Pi_{3/2}$, and $B^2\Sigma^{+}_{1/2}$ states. 
The dipole strength of the $A^2\Pi_{1/2}(000)\rightarrow X^2\Sigma^{+}_{1/2}(000)$ transition is therefore approximated as $|h^{\text{dip}}_{\text{AX}}|^2$. 
Using the dipole strength expressions in Eqns (\ref{ibm_dvc_A1}) to (\ref{ibm_dvc_B}), we calculate the ratio between the dipole strengths of a symmetry forbidden transition and the $A^2\Pi_{1/2}(000)\rightarrow X^2\Sigma^{+}_{1/2}(000)$ transition. 
The branching ratio of the symmetry-forbidden transition is then approximated as the product of this ratio of dipole strengths and the branching ratio of the $A^2\Pi_{1/2}(000)\rightarrow X^2\Sigma^{+}_{1/2}(000)$ transition obtained from the DVR calculations.


A comparison between branching ratios obtained from {\modColor the perturbation theory} and DVR calculations for CaOH, SrOH, and YbOH is summarized in Table \ref{ibm_table}.
In the DVR calculations, the calculated branching ratios can be divided into the contributions from the $A^2\Pi$ and $B^2\Sigma^{+}$ components.
In PT, the contribution from the $A^2\Pi_x$ and $A^2\Pi_y$ components to the branching ratio of the $A^2\Pi_{1/2}(000) \rightarrow X^2\Sigma^{+}_{1/2}(010)$ transition comes from the SOVC mechanism while the contribution from the $B^2\Sigma^{+}_{1/2}$ component comes from the DVC mechanism.
As a result, the ``SOVC'' and ``DVC'' contributions in PT are supposed to be comparable to the contributions from the ``$A^2\Pi$'' and ``$B^2\Sigma^{+}$'' components in the DVR calculations, respectively, for the $A^2\Pi_{1/2}(000) \rightarrow X^2\Sigma^{+}_{1/2}(010)$ and $A^2\Pi_{3/2}(000) \rightarrow X^2\Sigma^{+}_{1/2}(010)$ transitions.
Similarly, the ``DVC'' contribution in PT corresponds to the contribution from the ``$A^2\Pi$'' component in DVR for the $B^2\Sigma^{+}_{1/2}(000) \rightarrow X^2\Sigma^{+}_{1/2}(010)$ transition.

The calculated contributions to vibrational branching ratios obtained from DVR and PT agree are in general consistent with each other. The perturbative formulae give quantitatively correct prediction for these transitions except for the $A^2\Pi_{3/2}(000)\rightarrow X^2\Sigma^{+}_{1/2}(010)$ transition in YbOH, which may be tentatively attributed to the stronger spin-orbit mixing in YbOH.
The branching ratios of the $B^2\Sigma^{+}_{1/2}(000) \rightarrow X^2\Sigma^{+}_{1/2}(010)$ transition are one order of magnitude larger than those of the $A^2\Pi_{1/2}(000) \rightarrow X^2\Sigma^{+}_{1/2}(010)$ transition in CaOH, SrOH, and YbOH because of both smaller denominators in Eqn. (\ref{ibm_dvc_B}) and larger transition dipole moment.
These results agree well with the experimental measurements.

\subsubsection{TDMD mechanism}
We have calculated the adiabatic electronic transition dipole moment derivatives (TDMDs)
for the $A^2\Pi_{1/2}\rightarrow X^2\Sigma^{+}_{1/2}$, $A^2\Pi_{3/2}\rightarrow X^2\Sigma^{+}_{1/2}$, and $B^2\Sigma^{+}_{1/2}\rightarrow X^2\Sigma^{+}_{1/2}$ transitions
in the CaOH molecule at the SFX2C-1e-EOMEA-CCSD level by means of finite difference using a step size of $0.1$ unit of normal coordinate.
The adiabatic transition dipole moments have then been transformed to the quasidiabatic representation. 
The calculated quasidiabatic electronic transition dipole moment derivatives for CaOH molecule are summarized in Table \ref{TDMD}.
The calculated derivatives are in general small since the quasidiabatic wave functions vary slowly with respect to geometrical displacements by construction.

A comparison between the calculated branching ratios with and without electronic transition dipole moment derivatives is given in Table \ref{TDMD_br}.
The calculated derivatives of $h^{\text{dip}}_{\text{AX}}$ along the molecular bending modes are less than 0.0003 a.u..
As a result, the changes in the branching ratio of the $A^2\Pi_{1/2}(000)\rightarrow X^2\Sigma^{+}_{1/2}(010)$ transition due to the TDMD mechanism are negligible. 
On the other hand, the branching ratio of the $B^2\Sigma^{+}_{1/2}(000)\rightarrow X^2\Sigma^{+}_{1/2}(010)$ transition is enhanced by around 50\% since it has a larger transition dipole moment derivative of 0.0284 a.u..
The TDMD mechanism also slightly redistributes the branching ratios among the symmetry-allowed transitions. 
In the case of the CaOH molecule, it does not change the number of {\modColor the} transitions with branching ratios above $10^{-5}$. 
Based on these results, we may conclude that the TDMD mechanism is less significant than the DVC and SOVC mechanisms; the Franck-Condon approximation works well in the quasidiabatic representation.

\subsubsection{Fermi resonances}
\label{fermiResults}
The branching ratios of the transitions to near-degenerate vibrational states are redistributed due to the Fermi resonances. 
Consequently, the intensity of the $A^2\Pi_{1/2}(000) \rightarrow X^2\Sigma^{+}_{1/2}(020)$ transition in metal hydroxides
will be enhanced when the $X^2\Sigma^{+}_{1/2}(100)$ and $X^2\Sigma^{+}_{1/2}(020)$ states are energetically close to each other. 
Similar redistribution can also lead to non-negligible vibrational branching ratios for the $A^2\Pi_{1/2}(000)\rightarrow X^2\Sigma^{+}_{1/2}(120)$ and $A^2\Pi_{1/2}(000)\rightarrow X^2\Sigma^{+}_{1/2}(220)$ transitions.
The stronger Fermi resonances in CaOH compared to SrOH is an important reason why a nearly closed optical cycling for CaOH involves more transitions than for SrOH \cite{zhang2021CaSrYb}.

We have calculated the branching ratios of the $A^2\Pi_{1/2}(000)$ state for the deuterated species CaOD and SrOD as an example for the dependence of Fermi resonances to the vibrational energies. 
The results are summarized in Table \ref{H2D}.
The deuteration reduces the bending frequencies. 
In SrOD the deteuration reduces the $(010)$ frequency to 279 cm$^{-1}$,
which becomes close to half of the $(100)$ frequency value of 523 cm$^{-1}$.
This leads to significant coupling between (100) and (020) 
and hence a redistribution of the branching ratios among the transitions to these states. 
Consequently, the branching ratios of $A^2\Pi_{1/2}(000)\rightarrow X^2\Sigma^{+}_{1/2}(120)$ and $X^2\Sigma^{+}_{1/2}(040)$ transitions in SrOD are enhanced.
This renders SrOD less favorable for laser cooling than SrOH. 
In CaOD the difference between the $(100)$ and $(020)$ frequency values of 608 cm$^{-1}$ and 522 cm$^{-1}$ is similar to that in CaOH. 
The Fermi resonance contribution thus is also similar to that in CaOH.
As shown in Table \ref{H2D}, CaOD exhibits slightly lower branching ratios
for the transitions from $A^2\Pi_{1/2}(000)$ 
to $X^2\Sigma^{+}_{1/2}(020)$, $X^2\Sigma^{+}_{1/2}(120)$, and $X^2\Sigma^{+}_{1/2}(220)$. 
We may conclude that it is in general desirable to minimize the coupling between the (100) and (020) modes by increasing the energy gap between these states in search of candidate molecules for laser cooling.

\subsection{Benchmark studies of electron-correlation and basis-set effects}
\subsubsection{Core-correlation and basis-set effects on the transition properties}
Since the accuracy of the calculated branching ratios depends critically on that of the computed transition properties, we have performed benchmark calculations for the basis-set and core-correlation effects on these parameters. 
The transition properties have been calculated at the SFX2C-1e-EOMEA-CCSD level using the TZ and QZ basis sets for CaOH, SrOH, and YbOH.
The TZ and QZ basis sets recontracted for the SFX2C-1e scheme have been used in the calculations of LVC constants and transition dipole moments for SrOH and YbOH, while the uncontracted basis sets have been used in the other calculations.
All the electrons have been correlated in these calculations. To study the inner-shell correlation effects in YbOH, we have also carried out frozen-core calculations for YbOH with 24 core orbitals kept frozen in the CC calculations. 
The results are summarized in Table \ref{paraH}.

The QZ SOC matrix elements only differ from the TZ values by around 1\%.
The inner-shell correlation contributions in YbOH have the same magnitude, i.e., less than 1\% for transition dipole moments and about 2\% for the SOC matrix elements.
These contributions thus are insignificant.
On the other hand, the diabatic linear vibronic coupling constants $\lambda$ are more sensitive to both basis-set and core-correlation effects. 
The QZ LVC constant in CaOH is 20\% smaller than the TZ value.
Since the branching ratios for $\left|\Delta v_2\right| =1$ transitions are proportional to the square of the norm of $\lambda$, a 20\% reduction in $\lambda$ induces a more than 40\% decrease in the branching ratio for the $A^2\Pi_{1/2}(000) \rightarrow X^2\Sigma^{+}_{1/2}(010)$ transition.
The core-correlation contribution in YbOH is in the same direction as the basis-set effect. 
The complete basis set limit value for the LVC constant with all the electrons correlated 
is estimated to be around 100 cm$^{-1}$.

\subsubsection{The quality of the potential energy surfaces}
The accuracy of the calculated vibronic energy levels and branching ratios 
depends on the quality of the potential energy surfaces.
To estimate the remaining errors in the DVR results due to the errors of potential energy surfaces, we have compared the vibronic energy levels and branching ratios of the $A^2\Pi_{1/2}(000)\rightarrow X^2\Sigma^{+}_{1/2}(v_1 v_2 v_3)$ transitions in CaOH calculated using the EOMEA-CCSD/TZ, EOMEA-CCSD/QZ, and EOMEA-CCSDT/TZ potential energy surfaces in Tables \ref{vib ene} and \ref{br}.

The differences between the DVR vibrational energy levels obtained using these three variants of potential energy surfaces are in general small.
The $X^2\Sigma^{+}_{1/2}(010)$ frequency obtained using the EOMEA-CCSD/TZ PES differs by $-3$ cm$^{-1}$ and $6$ cm$^{-1}$ from 
those obtained from the EOMEA-CCSD/QZ and EOMEA-CCSDT/TZ PESs, respectively. 
The corresponding differences for the $X^2\Sigma^{+}_{1/2}(100)$ frequency are $-7$ cm$^{-1}$ and $2$ cm$^{-1}$. 
We note that the variations of the level position for the $A^2\Pi_{3/2}(100)$ state are larger than those of the other states.
A strong anharmonic mixing between the $A^2\Pi_{3/2}(100)$ state and the $A^2\Pi_{1/2}(020)$ state leads to unusual sensitivity to the quality of the potential energy surfaces.

As shown in Table \ref{br}, the residue basis-set effects and high-level correlation contributions are less significant for the branching ratios of the $A^2\Pi_{1/2}(000) \rightarrow X^2\Sigma^{+}_{1/2}(v_1 00)$ transitions with $v_1=1,2,3$.
In contrast, their contributions to the transitions involving the molecular bending modes are more substantial. 
The CCSD/QZ branching ratio for the $A^2\Pi_{1/2}(000) \rightarrow X^2\Sigma^{+}_{1/2}(02^00)$ transition is around 40\% smaller than the CCSD/TZ value. Furthermore, the triples correction is observed to enhance this transition by around 30\% when comparing the CCSD/TZ and CCSDT/TZ values.
More accurate PESs may be required to obtain a better agreement with the measured branching ratios for the $A^2\Pi_{1/2}(000) \rightarrow X^2\Sigma^{+}_{1/2}(02^00)$ and $A^2\Pi_{1/2}(000) \rightarrow X^2\Sigma^{+}_{1/2}(04^00)$ transitions.

The branching ratios for the symmetry-forbidden transitions, e.g., the $A^2\Pi_{1/2}(000)\rightarrow X^2\Sigma^{+}_{1/2}(010)$ transition, vary only slightly when using the different PESs.
The CCSD/TZ, CCSD/QZ, and CCSDT/TZ results in Table \ref{br}. 
This is consistent with the discussion of the intensity-borrowing mechanisms that the intensities for these transitions are dominated by the transition properties rather than the finer details of the PESs. 
On the other hand, the branching ratio for the $A^2\Pi_{1/2}(000)\rightarrow X^2\Sigma^{+}_{1/2}(010)$ transition obtained using the CCSD/TZ transition properties (the CCSD/TZ$^\ast$ results in Table \ref{br}) is larger by about 50\% than that obtained using the CCSD/QZ transition properties (the CCSD/TZ results). 
This difference in the branching ratio is consistent with the corresponding difference in
the linear vibronic coupling constant $\lambda$ as shown in Table \ref{paraH}.
Similarly, the energy splittings arising from SOC, i.e., the energy difference between $A^2\Pi_{1/2}(000)$ and $A^2\Pi_{3/2}(000)$ states, are dominated by the computed spin-orbit matrix elements and are relatively
insensitive to the basis-set and high-level-correlation effects on the potential energy surfaces.
We emphasize that the numbers of repumping lasers predicted using various PESs stay largely constant, demonstrating the robustness of the present computational scheme for the study of laser cooling of molecules.

\subsection{Prediction for the vibronic branching ratios of RaOH}
The RaOH molecule possesses a large effective electric field based on recent computational results \cite{Gaul2020,zhang2021calculations}, e.g.,
relativistic coupled-cluster calculations have given an effective electric field value of 55 GV/cm \cite{zhang2021calculations}.
{\modColor RaOH thus has high sensitivity for the precision measurement search of electron's electric dipole moment (eEDM).}
Furthermore, radium isotopes with non-zero nuclear spin, static octupole deformation, and low-lying nuclear states of opposite parity, such as $^{223}$Ra and $^{225}$Ra, have enhanced sensitivities to a nuclear Schiff moment~\cite{Flambaum2002NSMAtoms} by around three orders of magnitude compared to species with spherical nuclei \cite{Auerbach1996OctDev,Dobaczewski2005RaNSM}. 
The use of the parity doubling of the (010) vibrational level in RaOH, or other types of parity doublets in other radium-containing polyatomics, have experimental advantages due to high polarization in low-fields and methods to reject systematic errors \cite{YbOH2017,Yu2021RaOCH3}. 

Laser cooling can play an important role in enhancing the sensitivity of these measurements via the ability to access long coherence times and advanced quantum control,~\cite{YbOH2017} and
RaOH is expected~\cite{Isaev2017RaOH} to have similar electronic and vibrational structures as those in CaOH and SrOH, thus rendering it laser-coolable.  RaOH is an interesting species, as it is fairly unique in its combination of laser-coolability, parity doubling, high sensitivity to eEDM, nuclear structure enhancements, and relatively simple structure.

We have applied our computational scheme to predict the vibronic branching ratios for the $A^2\Pi_{1/2}(000)\rightarrow X^2\Sigma^{+}_{1/2}(v_1 v_2 v_3)$ optical cycling transition in RaOH to investigate the laser-coolability of this molecule.  The calculated linear vibronic coupling constant $\lambda$ as well as the spin-orbit coupling matrix elements $h^{\text{SO}}_\text{AA}$ and $h^{\text{SO}}_\text{AB}$ in RaOH amount to 11.9 cm$^{-1}$, 731.0 cm$^{-1}$, and 709.7 cm$^{-1}$, respectively. 
The calculated electronic transition dipole moments $h^{\text{dip}}_{\text{AX}}$ and $h^{\text{dip}}_{\text{BX}}$ are 2.6 a.u. and 2.1 a.u..
The calculated branching ratios above $10^{-5}$ for the $A^2\Pi_{1/2}(000)\rightarrow X^2\Sigma^{+}_{1/2}(v_1 v_2 v_3)$ transitions and the corresponding vibrational energy levels on the $X^2\Sigma^{+}_{1/2}$ state are summarized in Table \ref{raohBR}. 

The calculated Ra-O bond-length difference between the $X^2\Sigma^{+}_{1/2}$ and $A^2\Pi_{1/2}$ states at the SFX2C-1e-EOMEA-CCSD level is 0.012 {\AA}. 
It is smaller than the corresponding metal-oxygen bond-length differences in CaOH, SrOH, and YbOH, which amount to 0.021, 0.022, 0.034 {\AA}, respectively.
This small bond-length difference leads to more diagonal branching ratios for the Ra-O stretching mode, with a 99\% branching ratio for the $A^2\Pi_{1/2}(000)\rightarrow X^2\Sigma^{+}_{1/2}(000)$ transition.
Besides, the linear vibronic coupling constant $\lambda$ in RaOH is less than 12 cm$^{-1}$, one order of magnitude smaller than those in CaOH, SrOH, and YbOH.
As a result, except for a small branching ratio of 0.001\% for the $A^2\Pi_{1/2}(000) \rightarrow X^2\Sigma^{+}_{1/2}(010)$ transition, the nominally symmetry-forbidden transitions with odd values of $\Delta v_2$ possess negligible branching ratios. The branching ratios of RaOH are therefore calculated to be more favorable than CaOH or SrOH.
{\modColor  As shown in Table \ref{raohBR}, only seven transitions have branching ratios higher than 0.001\%. The sum of the branching ratios for the other transitions is lower than $10^{-5}$. }

Reliable computational results are valuable in determining whether the $A^{\prime 2}\Delta_{3/2}$ state is an intermediate state lying between the $A^2\Pi_{1/2}$ and $X^2\Sigma^{+}_{1/2}$, which is of critical importance to the laser cooling of RaOH.
We mention that the $A^{\prime 2}\Delta_{3/2}$ state is an intermediate state in YO \cite{Collopy2015}, while it has been shown to lie above the $A^2\Pi_{1/2}$ state in the case of RaF \cite{garcia2020spectroscopy,zaitsevskii2022accurate}.
We have carried out geometry optimizations for the $A^2\Pi_{1/2}$ and $A^{\prime 2}\Delta_{3/2}$ states of RaOH using X2CAMF coupled-cluster singles and doubles with a non-iterative triples [X2CAMF-CCSD(T)] method \cite{zhang22atomic,SOCC1liu2018} with the uncontracted ANO-RCC basis set for Ra and uncontracted cc-pVTZ basis sets for O and H.
The X2CAMF-CCSD(T) adiabatic excitation energy of the $A^{\prime 2}\Delta_{3/2}$ state is more than 1500 cm$^{-1}$ higher than that of the $A^2\Pi_{1/2}$ state. 
Based on the calculations of the energy differences using larger basis sets and the comparison between the CCSD and CCSD(T) results, we conclude that the estimated remaining errors from the residue basis-set effects and high-level correlation do not change the ordering of $A^{\prime 2}\Delta_{3/2}$ and $A^2\Pi_{1/2}$ states.
While the details of these calculations will be reported elsewhere, we emphasize here that there is no intermediate electronic state between $A^2\Pi_{1/2}$ and $X^2\Sigma^{+}_{1/2}$ in RaOH. 
{\modColor Optical cycling with the seven transitions given in Table \ref{raohBR} enables the scattering of more than 100,000 photons; based on a comparison to CaOH~\cite{vilas2022magneto},  taking into account the heavier mass and longer wavelength transitions of RaOH, this should be able to provide sufficient slowing and cooling for the molecules to be loaded into a 3D-MOT.}
RaOH thus is a very promising polyatomic radioactive molecule for laser cooling.

\section{Summary and outlook}
\label{summarySec}
We have analyzed the intensity-borrowing mechanisms for weak transitions in laser-coolable linear triatomic molecules, including the mechanisms involving non-adiabatic couplings, transition dipole moment derivatives, and Fermi resonances.  While these effects typically appear on the $\lesssim10^{-3}$ level, understanding them is critical for laser cooling.
Formulae to evaluate transition dipole strengths have been derived based on perturbation theory. The PT results for CaOH, SrOH, and YbOH are shown to agree well with those obtained from variational DVR calculations.
It has been shown that linear polyatomic molecules with small non-adiabatic coupling between the $A^2\Pi_{1/2}$ and $B^2\Sigma^{+}_{1/2}$ states and insignificant Fermi resonances are favorable candidates for laser cooling.

Benchmark calculations have been performed
to study the accuracy of the calculated transition properties and the vibronic branching ratios. 
The linear vibronic coupling constants are the most sensitive to basis-set and core-correlation effects among the transition properties. This leads to the sensitivity for the branching ratios of the symmetry-forbidden transitions as well. 
The high-level correlation and spin-orbit effects on the vibronic coupling constants will be studied in the future.

We have predicted the vibronic branching ratios for $A^2\Pi_{1/2}(000)\rightarrow X^2\Sigma^{+}_{1/2}(v_1 v_2 v_3)$ transitions in RaOH.
These transitions have quasidiagonal Franck-Condon factors with the $A^2\Pi_{1/2}(000)\rightarrow X^2\Sigma^{+}_{1/2}(000)$ transition accounting for 99\% of the transition intensities. Furthermore, because the $A^2\Pi_{1/2} - B^2\Sigma^{+}_{1/2}$ linear diabatic coupling constant in RaOH is an order of magnitude smaller than that in CaOH, most nominally forbidden transitions in RaOH have branching ratios negligible for laser cooling. 
With only {\modColor seven} transitions having branching ratios above $10^{-5}$, the RaOH molecule appears to be a promising candidate for laser cooling. 
Future work includes the calculations of accurate transition energies to facilitate the experimental preparation and detection of the RaOH molecule.


\section*{Supporting Information}

\begin{acknowledgement}

The authors thank John. F. Stanton (Gainesville) for stimulating discussions on the evaluation of transition dipole moments in diabatic representation , Peter Bryan Changala (Boston) for helpful discussions on the transition dipole moment derivatives mechanism, and Benjamin L. Augenbraun (Boston) for helpful discussions on intensity-borrowing mechanisms.
The work at the Johns Hopkins University has been supported by National Science Foundation under Grant No. PHY-2011794.
The work at California Institute of Technology has been supported by Heising-Simons Foundation award 2022-3361, Gordon and Betty Moore Foundation award GBMF7947, Alfred P. Sloan Foundation award G-2019-12502, and NSF CAREER award PHY-1847550. 
The computations at Johns Hopkins University were carried out at Advanced Research Computing at Hopkins (ARCH) core facility (rockfish.jhu.edu), which is supported by the National Science Foundation (NSF) under grant number OAC-1920103.  \\

\end{acknowledgement}

\clearpage
\begin{table}
  \begin{center}
  \caption{Branching ratios (in percentage) for the transitions repumped in the laser-cooling scheme for CaOH \cite{vilas2022magneto} from $A^2\Pi_{1/2}(000)$ to the vibrational levels of $X^2\Sigma^{+}$ with the harmonic approximation, single-state and multi-state DVR calculations, and experimental measurements. The harmonic calculations were performed using the bond lengths and harmonic frequencies obtained at the EOMEA-CCSD/QZ level.}
  \label{harmonic}
  \begin{tabular}{cccccccccc}
    \hline \hline
          Vibrational states     & Harmonic & Single-state & Multi-state{\modColor \cite{zhang2021CaSrYb}} & Exp.{\modColor \cite{zhang2021CaSrYb}}\\
    \hline
        $(000)$   & 95.439 & 95.518 & 95.429 & 94.59(29)      \\
        $(010)$   &  -     & -      &  0.063 & 0.099(6)    \\
        $(100)$   &  4.488 &  3.698 &  3.934 & 4.75(27)    \\
        $(02^00)$ &  0.013 &  0.596 &  0.298 & 0.270(17)    \\
        $(02^20)$ &  -     &  -     &  0.079 & 0.067(12)    \\
        $(110)$   &  -     &  -     &  0.003 & 0.0064(7)    \\
        $(200)$   &  0.060 &  0.138 &  0.157 & 0.174(16)   \\
        $(12^00)$ & <0.001 &  0.041 &  0.022 & 0.021(3)   \\
        $(12^20)$ &  -     &  -     &  0.005 & 0.005(1)   \\
        $(300)$   & <0.001 &  0.005 &  0.006 & 0.0068(8)  \\
    \hline \hline
  \end{tabular}
  \end{center}
\end{table}

\clearpage
\begin{table}
    \centering
    \caption{Contributions to the calculated branching ratios (in percentage) for the transitions from the $A^2\Pi_{1/2}(000)$, $A^2\Pi_{3/2}(000)$, and $B^2\Sigma^{+}_{1/2}(000)$ states to the $X^2\Sigma^{+}_{1/2}(010)$ state in CaOH, SrOH, and YbOH using {\modColor the perturbation theory (PT) and discrete variable representation (DVR) calculations. The $A^2\Pi$ and the $B^2\Sigma^{+}$ denote the contributions to the vibrational branching ratios from the $A^2\Pi$ and the $B^2\Sigma^{+}$ components in the DVR wave function.}}
    \begin{tabular}{cccccc}
    \hline\hline
       & \multirow{2}{*}{Initial state}   & \multicolumn{2}{c}{DVR} & \multicolumn{2}{c}{PT}\\
       & & $A^2\Pi$ & $B^2\Sigma^{+}$ & SOVC & DVC \\
         \hline
\multirow{3}{*}{CaOH} & $A^2\Pi_{1/2}(000)$   & 0.001 & 0.062 & 0.001 & 0.061 \\
                      & $A^2\Pi_{3/2}(000)$   & 0.001 & 0.066 & 0.001 & 0.064 \\
                      & $B^2\Sigma^{+}_{1/2}(000)$& 0.511 & 0.000 & - & 0.596 \\
        \hline
\multirow{3}{*}{SrOH} & $A^2\Pi_{1/2}(000)$   & 0.003 & 0.023 & 0.003 & 0.023 \\
                      & $A^2\Pi_{3/2}(000)$   & 0.055 & 0.035 & 0.051 & 0.029 \\
                      & $B^2\Sigma^{+}_{1/2}(000)$& 0.234 & 0.000 & - & 0.283 \\
        \hline
\multirow{3}{*}{YbOH} & $A^2\Pi_{1/2}(000)$   & 0.017 & 0.038 & 0.014 & 0.042 \\
                      & $A^2\Pi_{3/2}(000)$   & 0.073 & 0.045 & 0.084 & 0.095 \\
                      & $B^2\Sigma^{+}_{1/2}(000)$& 0.517 & 0.004 & - & 0.670 \\
    \hline\hline
    \end{tabular}
    \label{ibm_table}
\end{table}

\clearpage

\begin{table}
 \begin{center}
 \caption{Calculated quasidiabatic transition dipole moment derivatives (in a.u.) in CaOH. $Q_x$ and $Q_y$ denote the normal modes for the molecular bending along the $x$ and $y$ directions. 
 $Q_{\text{Ca-O}}$ and $Q_{\text{O-H}}$ denote the normal modes for Ca-O and O-H stretching.
 }
 \label{TDMD}
 \begin{tabular}{ccccc}
    \hline \hline
    transition dipole & $\partial/\partial Q_x$ & $\partial/\partial Q_y$ & $\partial/\partial Q_{\text{Ca-O}}$ & $\partial/\partial Q_{\text{O-H}}$\\
    \hline
    $\langle X^2\Sigma^{+}| \mu_x | A^2\Pi_x \rangle$ & 0.0  & 0.0 & 0.0428 & 0.0002 \\
    $\langle X^2\Sigma^{+}| \mu_y | A^2\Pi_x \rangle$ & 0.0 & 0.0 & 0.0 & 0.0  \\
    $\langle X^2\Sigma^{+}| \mu_z | A^2\Pi_x \rangle$ & -0.0003 & 0.0 & 0.0 & 0.0  \\
    \hline
    $\langle X^2\Sigma^{+}| \mu_x | A^2\Pi_y \rangle$ & 0.0  & 0.0 & 0.0 & 0.0 \\
    $\langle X^2\Sigma^{+}| \mu_y | A^2\Pi_y \rangle$ & 0.0 & 0.0 & 0.0428 & 0.0002  \\
    $\langle X^2\Sigma^{+}| \mu_z | A^2\Pi_y \rangle$ & 0.0 & -0.0003 & 0.0 & 0.0  \\
    \hline
    $\langle X^2\Sigma^{+}| \mu_x | B^2\Sigma^{+} \rangle$ & 0.0284  & 0.0 & 0.0 & 0.0 \\
    $\langle X^2\Sigma^{+}| \mu_y | B^2\Sigma^{+} \rangle$ & 0.0 & 0.0284 & 0.0 & 0.0  \\
    $\langle X^2\Sigma^{+}| \mu_z | B^2\Sigma^{+} \rangle$ & 0.0 & 0.0 & 0.0435 & 0.0004  \\
    \hline \hline
 \end{tabular}
 \end{center}
\end{table}

\clearpage

\begin{table}
 \begin{center}
 \caption{Calculated branching ratios (in percentage) for the transitions from $A^2\Pi_{1/2}(000)$ and $B^2\Sigma^{+}_{1/2}(000)$ to the vibrational levels of $X^2\Sigma^{+}$ in CaOH with or without the inclusion of the transition dipole moments derivatives (TDMDs).}
 \label{TDMD_br}
 \begin{tabular}{ccccc}
    \hline \hline
    Initial state & \multicolumn{2}{c}{$A^2\Pi_{1/2}(000)$} & \multicolumn{2}{c}{$B^2\Sigma^{+}_{1/2}(000)$}\\
       &  without TDMDs & with TDMDs &  without TDMDs & with TDMDs \\
    \hline
        $(000)$   & 95.429 & 94.906 & 97.259 & 96.544 \\
        $(010)$   &  0.063 &  0.062 &  0.511 &  0.730 \\
        $(100)$   &  3.934 &  4.389 &  2.082 &  2.528 \\
        $(02^00)$ &  0.298 &  0.330 &  0.033 &  0.049 \\
        $(02^20)$ &  0.079 &  0.077 & <0.001 & <0.001 \\
        $(110)$   &  0.003 &  0.003 &  0.012 &  0.021 \\
        $(03^10)$ & <0.001 & <0.001 &  0.001 &  0.001 \\
        $(200)$   &  0.157 &  0.188 &  0.092 &  0.116 \\
        $(12^00)$ &  0.022 &  0.027 &  0.001 &  0.001 \\
        $(12^20)$ &  0.005 &  0.006 & <0.001 & <0.001 \\
        $(210)$   & <0.001 & <0.001 &  0.001 &  0.002 \\
        $(050)$   & <0.001 & <0.001 &  0.001 &  0.001 \\
        $(300)$   &  0.006 &  0.008 &  0.004 &  0.006 \\
        $(22^00)$ &  0.002 &  0.002 & <0.001 & <0.001 \\
    \hline \hline
 \end{tabular}
 \end{center}
\end{table}

\clearpage

\begin{table}
 \begin{center}
 \caption{Multi-state DVR branching ratios (in percentage) for the transitions from $A^2\Pi_{1/2}(000)$ to the vibrational levels of $X^2\Sigma^{+}$ in CaOD and SrOD using the EOMEA-CCSD/QZ potential energy surfaces and transition properties.}
 \label{H2D}
 \begin{tabular}{ccc}
    \hline \hline
    Vibrational state & CaOD & SrOD\\
    \hline
    $(000)$           & 95.292  & 94.626 \\
    $(010)$           & 0.083   &  0.038 \\
    $(100)$           & 4.304   &  4.700 \\
    $(02^00)$         & 0.043   &  0.380 \\
    $(02^20)$         & 0.076   &  0.032 \\
    $(110)$           & 0.003   &  0.001 \\
    $(200)$           & 0.178   &  0.129 \\
    $(12^00)$         & 0.008   &  0.054 \\
    $(12^20)$         & 0.005   &  0.001 \\
    $(03^10)$         & $<$0.001&  0.001 \\
    $(04^00)$         & $<$0.001&  0.028 \\
    $(04^20)$         & $<$0.001&  0.002 \\
    $(300)$           & 0.007   &  0.005 \\
    $\text{other states}$            & 0.001   &  0.003 \\
    \hline \hline
 \end{tabular}
 \end{center}
\end{table}

\clearpage
\begin{table}
  \begin{center}
    \caption{Calculated transition properties at the SFX2C-1e-EOMEA-CCSD level. The numbers in the parentheses denote the numbers of frozen-core orbitals in coupled-cluster calculations. The other calculations have correlated all the electrons.}
    \label{paraH}
    \begin{tabular}{ccccccccccc}
      \hline\hline
       & ~ & \multicolumn{2}{c}{CaOH}  & ~ & \multicolumn{2}{c}{SrOH} & ~ &  \multicolumn{3}{c}{YbOH}\\
        & ~ & TZ & QZ  & ~ & TZ & QZ & ~ & TZ (24) & QZ (24) & TZ \\
      \hline
      $\lambda$/cm$^{-1}$                 & ~ & 146.5 & 121.8 & ~ & 89.4  & 70.3  & ~ & 139.7  & 124.8 & 127.3  \\
      $h^{\text{SO}}_\text{AA}$/cm$^{-1}$ & ~ & 32.9  & 33.2  & ~ & 121.0 & 122.2 & ~ & 486.8 & 482.9 & 497.4   \\ 
      $h^{\text{SO}}_\text{AB}$/cm$^{-1}$ & ~ & 25.5  & 26.4  & ~ & 103.9 & 106.1 & ~ & 415.5 & 411.7 & 426.1   \\ 
      $h^{\text{dip}}_{\text{AX}}$/a.u.   & ~ & 2.32  & 2.30  & ~ & 2.52  & 2.49  & ~ & 2.30 & 2.28 & 2.32   \\ 
      $h^{\text{dip}}_{\text{BX}}$/a.u.   & ~ & 1.86  & 1.85  & ~ & 2.04  & 2.02  & ~ & 1.86 & 1.85 & 1.87   \\
      \hline \hline
    \end{tabular}
  \end{center}
\end{table}

\clearpage
\begin{table}
  \begin{center}
  \caption{Calculated vibrational energy levels (in cm$^{-1}$) for the $X^2\Sigma^{+}_{1/2}$, $A^2\Pi_{1/2}$, and $A^2\Pi_{3/2}$ states in CaOH using a variety of EOMEA-CC potential energy surfaces. ``CCSD'' and ``CCSDT'' in this Table stand for EOMEA-CCSD and EOMEA-CCSDT, respectively.}
  \label{vib ene}
  \begin{tabular}{ccccccc}
    \hline \hline
    Vibrational state &  CCSD/QZ{\modColor \cite{zhang2021CaSrYb}} &  CCSD/TZ &  CCSDT/TZ &  Experiments{\modColor \cite{zhang2021CaSrYb}} \\
    \hline
    $X^2\Sigma^{+}_{1/2}(000)$   & 0     & 0     & 0     & 0     \\
    $X^2\Sigma^{+}_{1/2}(010)$   & 355   & 352   & 346   & 352.9   \\
    $X^2\Sigma^{+}_{1/2}(100)$   & 611   & 604   & 602   & 609.0   \\
    $X^2\Sigma^{+}_{1/2}(02^00)$ & 695   & 686   & 674   & 688.7   \\
    $X^2\Sigma^{+}_{1/2}(02^20)$ & 718   & 710   & 697   & 713.0   \\
    $X^2\Sigma^{+}_{1/2}(110)$   & 955   & 944   & 935   & 952   \\
    $X^2\Sigma^{+}_{1/2}(03^10)$ & 1045  & 1030  & 1013  & 1018  \\
    $X^2\Sigma^{+}_{1/2}(03^30)$ & 1091  & 1077  & 1058  & -     \\
    $X^2\Sigma^{+}_{1/2}(200)$   & 1215  & 1202  & 1196  & 1210.2  \\
    $X^2\Sigma^{+}_{1/2}(12^00)$ & 1293  & 1277  & 1264  & 1283  \\
    $X^2\Sigma^{+}_{1/2}(12^20)$ & 1310  & 1294  & 1279  & 1302  \\
    $X^2\Sigma^{+}_{1/2}(04^00)$ & 1388  & 1368  & 1346  & -     \\
    $X^2\Sigma^{+}_{1/2}(04^20)$ & 1410  & 1390  & 1367  & -     \\
    $X^2\Sigma^{+}_{1/2}(210)$   & 1548  & 1531  & 1518  & -     \\
    $X^2\Sigma^{+}_{1/2}(300)$   & 1815  & 1798  & 1788  & -     \\
    $X^2\Sigma^{+}_{1/2}(22^00)$ & 1890  & 1869  & 1856  & -     \\
    \hline         
    $A^2\Pi_{1/2}(000)$      & 0     & 0     & 0     & 0  \\
    $A^2\Pi_{3/2}(000)$      & 67    & 67    & 67    & 66.8  \\
    $A^2\Pi_{1/2}(010)$      & 346   & 337   & 331   & 345  \\
    $A^2\Pi_{1/2}(010)$      & 362   & 355   & 349   & 360  \\
    $A^2\Pi_{3/2}(010)$      & 429   & 421   & 415   & 425  \\
    $A^2\Pi_{3/2}(010)$      & 446   & 442   & 435   & 445  \\
    $A^2\Pi_{1/2}(100)$      & 623   & 618   & 613   & 628.7  \\
    $A^2\Pi_{3/2}(100)$      & 682   & 704   & 699   & 695.9  \\
    \hline \hline
  \end{tabular}
  \end{center}
\end{table}

\clearpage
\begin{table}
  \begin{center}
  \caption{Calculated branching ratios (in percentage) for the transitions from $A^2\Pi_{1/2}(000)$ to the vibrational levels of $X^2\Sigma^{+}$ in CaOH using a variety of EOMEA-CC potential energy surfaces. All the calculations have been carried out using the EOM-CCSD/QZ transition properties, except that the ``CCSD/TZ$^{\star}$'' column has used the EOM-CCSD/{\modColor TZ} transition properties.}
  \label{br}
  \begin{tabular}{cccccccc}
    \hline \hline
          Vibrational states     & CCSD/QZ{\modColor \cite{zhang2021CaSrYb}} & CCSD/TZ & CCSD/TZ$^{\star}$ & CCSDT/TZ & Exp.{\modColor \cite{zhang2021CaSrYb}} \\
    \hline
        $(000)$   & 95.429 & 94.392 & 94.371 & 94.482 & 94.59(29)     \\
        $(010)$   &  0.063 &  0.053 &  0.076 &  0.060 & 0.099(6)   \\
        $(100)$   &  3.934 &  4.498 &  4.702 &  4.464 & 4.75(27)   \\
        $(02^00)$ &  0.298 &  0.491 &  0.487 &  0.633 & 0.270(17)   \\
        $(02^20)$ &  0.079 &  0.099 &  0.097 &  0.101 & 0.067(12)   \\
        $(110)$   &  0.003 &  0.003 &  0.004 &  0.003 & 0.0064(7)   \\
        $(03^10)$ & <0.001 & <0.001 &  0.001 &  0.001 & 0.0034(8)  \\
        $(200)$   &  0.157 &  0.201 &  0.201 &  0.181 & 0.174(16)  \\
        $(12^00)$ &  0.022 &  0.041 &  0.041 &  0.054 & 0.021(3)  \\
        $(12^20)$ &  0.005 &  0.008 &  0.007 &  0.007 & 0.005(1)  \\
        $(04^00)$ & <0.001 &  0.001 &  0.001 &  0.001 & 0.0021(7) \\
        $(04^20)$ & <0.001 &  0.001 &  0.001 &  0.001 & -         \\
        $(300)$   &  0.006 &  0.008 &  0.008 &  0.007 & 0.0068(8) \\
        $(22^00)$ &  0.002 &  0.003 &  0.003 &  0.004 & 0.0020(7) \\
    \hline \hline
  \end{tabular}
  \end{center}
\end{table}

\clearpage
\begin{table}
  \begin{center}
  \caption{Calculated vibrational energy levels (in cm$^{-1}$) on the $X^2\Sigma^{+}_{1/2}$ state of RaOH and the corresponding branching ratios (in percentage) above $10^{-5}$ for the transitions from $A^2\Pi_{1/2}(000)$.}
  \label{raohBR}
  \begin{tabular}{ccc}
    \hline \hline
          Vibrational states & energy level & branching ratio  \\
    \hline
        $(000)$   &  0   & 98.972    \\
        $(010)$   &  337 &  0.001 \\
        $(100)$   &  475 &  0.863 \\
        $(02^00)$ &  646 &  0.138  \\
        $(02^20)$ &  678 &  0.007  \\
        $(200)$   &  947 &  0.012 \\
        $(12^00)$ & 1111 &  0.005 \\
    \hline \hline
  \end{tabular}
  \end{center}
\end{table}

\clearpage

\begin{tocentry}
    \begin{center}
    \includegraphics[width=6.5cm]{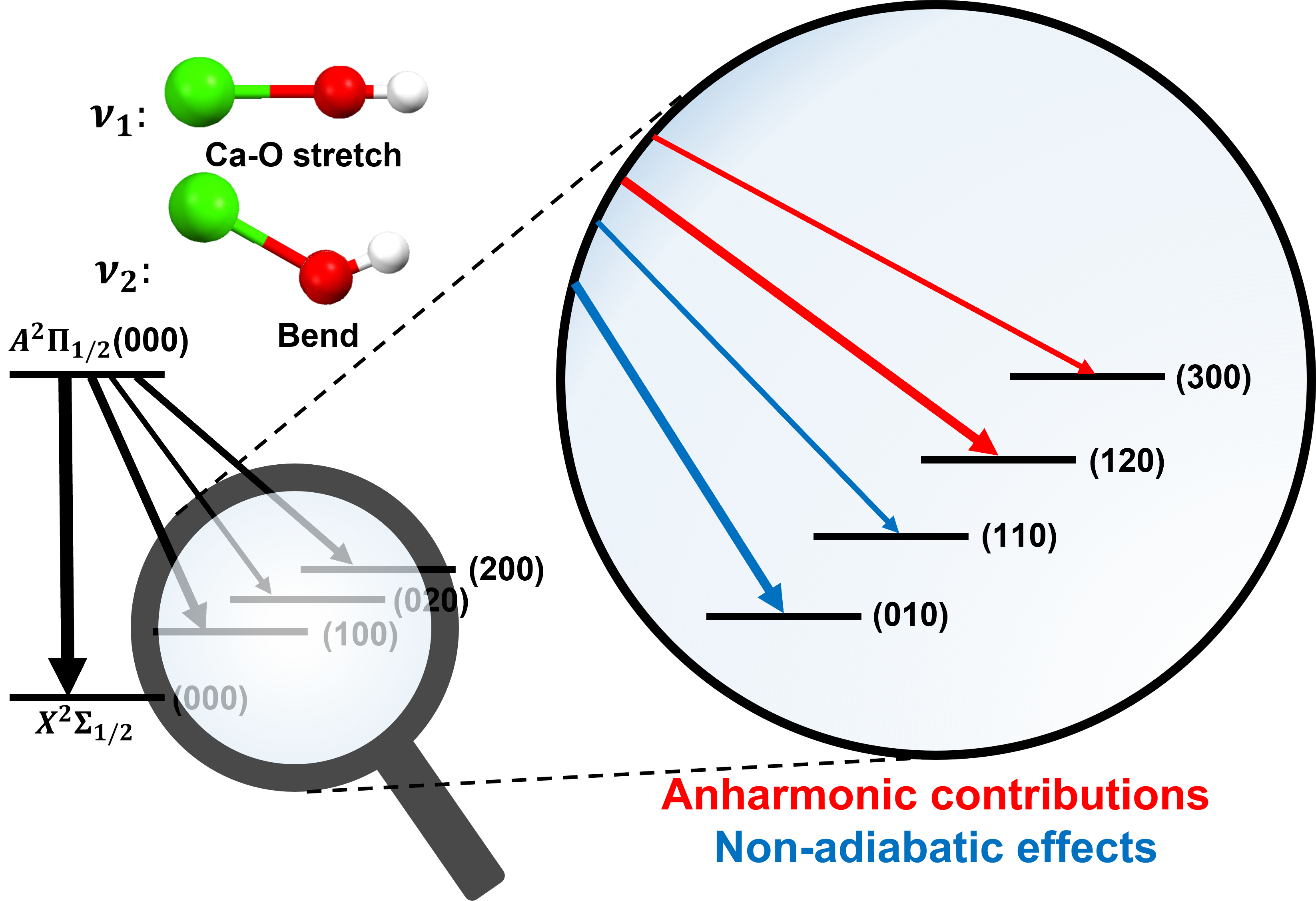}
    \end{center}
\end{tocentry}



\bibliography{reference}

\end{document}